\documentclass{llncs}
\usepackage{a4,graphicx,fancyhdr, vmargin}
\usepackage{url}
 
\begin{document}
\title{Utility and Disclosure Risk for Differentially Private Synthetic Categorical  Data}

  \author{Gillian M Raab\inst{1}}
  \institute{Scottish Centre for Administrative Data Research, University of Edinburgh, Scotlnd, UK, gillian.raab@ed.ac.uk}
\maketitle
\begin{abstract}
This paper introduces two methods of creating differentially private (DP) synthetic data that are now incorporated into the \textit{synth}\textbf{\textit{pop}} package for \textbf{R}. Both are suitable for synthesising categorical data, or numeric data grouped into categories. Ten data sets with varying characteristics were used to evaluate the methods. Measures of disclosiveness and of utility were defined  and calculated.  The first method is to add DP noise to a cross tabulation of all the variables and create synthetic data by a multinomial sample from the resulting probabilities. While this method certainly reduced disclosure risk, it did not provide synthetic data of adequate quality for any of the data sets. The other method is to create a set of noisy marginal distributions that are made to agree with each other with an iterative proportional fitting algorithm and then to use the fitted probabilities as above. This proved to provide useable synthetic data for most of these data sets at values of the differentially privacy parameter $\epsilon$ as low as 0.5. The relationship between the disclosure risk and $\epsilon$ is illustrated for each of the data sets. Results show how the trade-off between disclosiveness and data utility depend on the characteristics of the data sets. 
\end{abstract}

\section{Introduction}
Differential privacy (DP)\footnote{This acronym will also be used for ``differentially private".} \cite{Dwork2006a} is considered by theoretical computer scientists to be the most rigorous system of protecting the privacy of individuals in data released to the public. Formally, the release of a statistic that has been altered to comply with $\epsilon$-DP limits, restricts the absolute value of the log-likelihood-ratio of obtaining this result with the complete data to the that from data without any one individual. Small values of $\epsilon$ lead to greater protection of privacy by increasing the distance between the original and the $\epsilon$-DP result, while large values do little to preserve privacy and give results closer to those that would be found from the original data. The initial development of DP
 focussed on privacy without reference to utility, although it was recognised that DP results could provide answers far from the original, expecially for statistics based on small data sets. 

In contrast, the generation of synthetic data, as initially implemented in our \textit{synth}\textbf{\textit{pop}} package for \textbf{R} \cite{synthpop_last_CRAN}, provides a tool to obtain and evaluate the best possible utility but without any formal privacy  guarantee, although the fact that no synthetic record corresponds to a real individual gives some reassurance. Additional privacy protection for the output from the \textit{synth}\textbf{\textit{pop}}	 package is afforded by the use of statistical disclosure control measures (sdc) that are available as part of the package. This includes the removal of ``replicated uniques" defined as records that are unique in the synthetic data, but are also present and unique in the original data, as well as other methods such as smoothing and top and bottom coding of numeric data.

The original conception of DP was that a DP mechanism would be designed to answer individual queries, or a series of queries.  When a series of queries are answered for the same data set, the person receiving these DP answers would have 
increased the disclosure risk to the sum of all the $\epsilon$s in the individual queries. This appeared to rule out the possibility of using it to generate synthetic data for which there would be no limit on the number of queries. The few early attempts to create DP synthetic data seemed to result in poor utility, but the last ten years have seen many developments of practical DP methodology, much of it encouraged by endeavours to apply DP to outputs from the 2020 US Census \cite{abowd2018} \cite{handbook} \cite{Hawes2020} \cite{Garfinkel}. These initiatives are not universally accepted as a good thing. There have been criticisms  that DP distorts tasks such as redistricting voter areas \cite{kenny} as well as claims that DP does not prevent the leakage of confidential information \cite{useless}. 

The computer science literature uses different language and conventions from the statistical literature. This makes it difficult for statisticians to evaluate, especially as some of the DP algorithms are complex, see e.g. \cite{Hay2016}.
In this paper we introduce two easy-to-understand models for creating synthetic data for grouped data that are now incorporated into the development version of \textit{synth}\textbf{\textit{pop}}, currently on github \footnote{It can be installed from \url{https://github.com/bnowok/synthpop}}. 
Since its introduction there have been several modifications of the original DP definition, such as $\epsilon$-$\delta$-DP and other variants \cite{variants}. In this paper we use only the original definition and use only the Laplace methanism for adding DP noise. 
The models are evaluated on 10 data sets with different characteristics, and the utility and disclosure risk of the synthetic data are assessed. These methods could be developed further and they may not work as well as those developed by teams working with the US Census, but they can  provide an experience of creating DP synthetic data to a wider group of people who can gain experience with the method.

\section{Methods for creating synthetic data sets}
\subsection{Without DP guarantee}
Many methods of creating synthetic data are based on original proposals in  1993 \cite{little_1993} \cite{rubin_1993} that began to be implemented and developed from 2003 \cite{rrr_2003}. There is already a large literature for creating completely synthetic data, without reference to DP; see Drechsler\'s monograph \cite{Drechsler_book} for a comprehensive review.
While the original developments in this field built on ideas from multiple imputation, it is simpler to consider the methodology as generating data from a model fitted to the real data \cite{raab_jpc}.  Synthetic data are simulated from this model using either the fitted parameters, or a sample or samples from the posterior distribution of the fitted parameters. The parameters may be obtained by a fit to the joint distribution of all the variables or, more commonly, by defining the joint distribution in terms of a series of conditional distributions. 
\subsubsection{Synthesis methods implemented in \textit{synth}\textbf{\textit{pop}}}
 Synthesising from conditional distributions was the only method available in the original version of the \textit{synth}\textbf{\textit{pop}} package. It provides great flexibility by allowing a choice of model for each conditional distribution, by changing the order of the conditional distributions and by defining the predictors to be used for each conditional distribution. This allows the synthesiser to improve the quality of the synthetic data using tools to evaluate its utility\cite{utility2022} that are now part of the package. 

From Version 1.5 \cite{synthpop_5.0-0_CRAN} in 2018, \textit{synth}\textbf{\textit{pop}} includes two methods based on a log-linear fit to the joint distribution for data when all variables are categorical. The first ($catall$) fits a saturated model by selecting a sample  from a multinomial distribution with probabilities calculated from the complete cross-tabulation of all the variables in the data set. This is very close to a method recently  proposed by Jackson et al. \cite{jackson_et_al2021}. Jackson et al. also use a saturated model, but they generate data from a Poisson distribution. They present interesting results of different ways in which the synthetic data can be made less disclosive by simulating from overdispersed distributions, such as the negative binomial or the Poisson inverse Gamma.  The \textit{catall} procedure can be made exactly equivalent to the Poisson method if the number of records in the synthetic data is itself sampled from a Poisson distribution. When the cross-tabulation contains cells with zero counts these will be reproduced as zero cells in the synthetic data unless a small positive quantity ($\alpha$) is added to each cell in the table. The \textit{synth}\textbf{\textit{pop}} package allows for this by spreading a total count defined by the parameter $nprior = \alpha/k$, where $k$ is the number of cells in the table, evenly across all the cells in the table that are not structural zeros (e.g. a cell for the type of qualification for people with no qualifications). The second method \textit{ipf} fits log-linear models to a set of margins defined by the user using the method of iterative proportional fitting\footnote{also known as the RAS algorithm and as raking in Computer Science.} implemented in the package \textit{mipfp} in \textbf{R}. As for \textit{catall} the parameter $\alpha$ can be set to allow non-structural zeros to appear in the synthetic data. Both \textit{catall} and \textit{ipf} are only feasible for a small number of variables because of the need to create very large cross-tabulations of all the variables. \textit{synth}\textbf{\textit{pop}} currently prints a warning if the total cells exceed $10^8$, though a user can attempt to increase this. 

\subsubsection{Other methods}
More recently, the computer science and machine learning literature has introduced many new methods for creating synthetic data. A web search for ``synthetic data  solutions" generates links to many firms offering synthetic data services\footnote{Hazy \url{https://hazy.com/}, Accessed 18 May 2022.} \footnote{Mostly AI  \url{https://mostly.ai/ebook/synthetic-data-for-enterprises}, Accessed 18 May 2022}. One can even find a link to a list of firms  who can provide synthetic data for businesses\footnote{ \url{https://research.aimultiple.com/synthetic-data/},Multiple AI: In-Depth Synthetic Data Guide, Accessed 18 May 2022.}. Many of the machine learning algorithms are based on Generalized Adversarial Networks (GANs)\cite{goodfellow}. These methods involve an iterative process where a 'Generator' fits a model to the data and a 'Discriminator' attempts to distinguish between the real data and the model. Feedback to the 'Generator' is used to improve the fit of the data at the next iteration.
		 
\subsubsection{Compatible generative models}
Users of synthetic data can use it to calculate any summary statistic or to estimate any statistical model. The results obtained and the standard errors calculated for them \cite{raab_jpc} will only be unbiassed if the  model used in generating the synthetic data is at least as complex as that used in the analysis. We will refer to this aspect as the use of a generative model that is compatible with the analysis. Even more important is the condition that the original data is well represented by the generative model. The synthetic data may then appear to provide a good fit for  an analysis compatible with the synthesis model, when an analysis of the original data would have indicated a lack-of-fit.  If the analysis being used is  incompatible with the generative model creating the synthetic data, results may be biassed compared to those from the original data; i.e. synthetic data will have poor utility. Synthesis methods that appear to provide the most useful synthetic data are adaptive methods where the fitting procedure explores the relationships between variables to obtain the best fit. Such methods include a full conditional model where all conditional distributions are fitted by an adaptive CART model (the default model in \textit{synth}\textbf{\textit{pop}}) as well as the many different implementations of GANs.

The method \textit{catall} is compatible with any analysis of categorical data. Synthesis with \textit{ipf} is only compatible with analyses for which the sufficient statistics are members of the set of margins used to contrain the iterative fitting. If the data have significant interactions that are not included in the set of margins, the results of any incompatible analyses will be biassed. Poor utility will be found in these cases when evaluated from a fit to an incompatible model. However, unlike synthetic data created by adaptive methods, knowledge of the marginals will allow the user of synthetic data to know which analyses can be trusted. For example, one could create synthetic data with  \textit{ipf} with all two-way marginals and their interactions with the outcome of interest. Then a logistic regression of the outcome on all other variables and their two-way interactions would be a compatible analysis model.

\subsection{Adapting methods for synthetic data to make them DP}
\subsubsection{Background}
For a comprehensive review of practical methods for DP data synthesis see Bowen and Liu,  2020 \cite{BowenLiu2020}. This section will provide a brief summary, focussing mainly on the methods for categorical data used in this paper.

The earliest real application of DP synthetic data was the data set that sits behind the US Census Bureau's online tool 'On the Map'\footnote{see \url{https://lehd.ces.census.gov/applications/}, Acessed 16th March 2022.} that allows visualisation of where people work and where they live. The technique used was to add noise to the raw data for this example by either simulating from data with a Multinomial-Dirchilet distribution or by adding Laplace noise to the counts \cite{aandvilh} \cite{onthemap}. The values of $\epsilon$ had to be larger than was desirable and data had to be modified to make it appear plausible.

A more promising approach to creating synthetic data is to add DP noise to the parameters of the joint distribution, rather than to the raw data. The $\epsilon$ for the whole dataset is the sum of the individual values for each parameter. Shlomo \cite{shlomo} has recently illustrated this approach for synthesising data with a multivariate Normal distribution.

\subsubsection{Methods used in the NIST callenge} 
In recent years several groups have developed methods and software for creating DP synthetic data. These initiatives have been encouraged by a series of challenges, with substantial prizes, promoted by the National Institute of Standards and Technology (NIST)\footnote{https://www.nist.gov/ctl/pscr/open-innovation-prize-challenges/past-prize-challenges, where you will find details of the winning methods and links to some of the software used.}. Participating teams had to provide synthetic versions of the data for each challenge where the versions satisfied DP for a given set of values of $\epsilon$ and $\delta$. The teams had to provide code that would convince the judges that their data satisfied $\epsilon$-$\delta$ differential privacy. The software used by many of the teams who entered can be accessed via the NIST web site. Those submissions that passed this hurdle were then compared using utility metrics. Bowen and Snoke \cite{BowenSnoke2021} have published a detailed evaluation of the synthetic data that were submitted and describe the methods used by the teams. 

Two main types of methods were employed in the challenge. The first  was those based on DP GANs. GANS can be easily made DP by providing feedback from the Discriminator to the Generator via the results of DP queries. The total $\epsilon$ for the method is the sum of those used at each iteration and the iterations have to stop once this DP budget is exhausted. The second method, used by many teams, was to generate synthetic data from a model fitted to the margins of the data. Each margin is first made DP and a model is fitted that assumes that only the interactions defined by the chosen margins are present in the population from which the data can be considered to have been generated.  After noise has been added to the margins to make them DP, they no longer sum to the same totals for lower order marginals. Several methods have been used to make the margins agree and ensure they are non-negative. Once this is achieved synthetic data can be generated from a model fitted to the DP margins. The resulting synthetic data are DP because once the DP margins are created, any data derived from them without further input from the original data will also be DP\footnote{This is described as robustness to post-processing} . The choice of margins cannot be based on an analysis of the data to be synthesised as this would incur an addition to the privacy budget. The NIST teams used analyses of other similar data sets to inform their choices or, in some cases, used part of their privacy budget to identify the margins to fit. In the evaluation the teams who used  marginal methods scored more highly than those using GANs. One such team who scored highly on all challenges \cite{mcKwinning} has provided a detailed description of their methodology including their model choice and the post-processing methods used.

\subsubsection{DP methods in \textit{syn}\textbf{pop}}
The methods $catall$ and $ipf$ can be made DP. The first by a method similar to the method used by Abowd and Villhuber\cite{aandvilh} in 2008. The second is similar to the marginal models used by the NIST teams. Note that the choice of margins $M$ to use for DP $ipf$ cannot be made from an analysis of the original data, unless the analyses to determine $M$ contributes to the privacy budget. In this preliminary investigation $M$ consists of all two-way interactions in all cases.   The process of making these two methods DP, as implemented in the latest version  \textit{synth}\textbf{\textit{pop}}\footnote{See footnote 2. It will be made available on CRAN after Version 1.7.0}, involves the following steps:
\begin{enumerate}{
	\item{Determine the value of $\epsilon$ to use.}
	\item{Create  cross-tabulations of all the variables (\textit{catall}) or of the selected margins (\textit{ipf}).}
	\item{If the parameter $nprior > 0$  add $nprior/n_{cells}$ to each cell in every table, where $n_{cells}$ is the number of cells over which the prior is to be spread.}
	\item {Add Laplace noise with dispersion  parameter 1/$\epsilon$ 
	(\textit{catall}) or $M/\epsilon$ for (\textit{ipf}), where $M$ is the number of margins fitted to each cell of the table.}
	\item{Set any negative counts to zero or a small positive value.}
	\item{Rescale the counts to become probabilities that sum to unity.}
\item{For \textit{catall} create the synthetic data as a multinomial distribution from this probability vector with the selected sample size.}
 \item{For \textit{ipf} use the $Ipfp$ algorithm from the $mipfp$ package\footnote{see \url{https://CRAN.R-project.org/package=mipfp}} for \textbf{R} to obtain a fit to the 
 probabilities calculated from the noisy marginals, and then generate synthetic data as for \textit{catall} }
}
\end{enumerate}

Note that when iterative proportional fitting is applied to the probabilities from incompatible margins, rather than compatible counts, the margins are adjusted as part of the process so they become compatible with each other.  Convergence for iterative proportional fitting is always slow compared to the Newton-Raphson algorithms, but it is usually slow but sure. It becomes even slower when the initial margins are incompatible, and in a few cases it may fail to converge in a large number of iterations; see Table 3 for details. For the case when the margins are compatible the eventual  convergence is guaranteed, but is not clear if this is the case when margins are not compatible.  Steps 5 and 6 are a post-processing steps that can have a considerable influence on disclosiveness and utility.

\section{Measures of utility and disclosure risk for synthetic categorical data}

\subsection{Disclosure risk}
For DP synthetic data $\epsilon$ provides a measure of disclosure risk. It is desirable to have another measure of disclosure risk that can be calculated for non-DP synthetic data and that can be compared to $\epsilon$ for DP synthetic data. One such measure is the percentage of  unique records in the synthetic data that are also unique in the original data, designated as $ru$ (replicated uniques)\footnote{also sometimes termed ``correct matches''}. Below we introduce the notation used and define replicated uniques for the case when all variables are categorical.  The measure $ru$ will depend on the disclosiveness of the original data. In the extreme, if there are no unique records in the original data then $ru$ will always be zero. This measure relates to the expected behaviour of an intruder who knows the characteristics of a unique individual in the data base and attempts to identify them. A more nuanced version of this type of measure has been discussed by Taub et al.\cite{taub}, but this relates to a particular  target variable for which the value might be determined from a set of key variables. To obtain a measure for a data set this would need to be averaged over a selection of targets  Replicated uniques were used by Jackson et al.\cite{jackson_et_al2021} in their evaluation of syntheses from saturated models.

\begin{table}[ht]
	\centering
	\begin{tabular}{lll}
	$N$	& Number of observations in the original data \\
	$k$ &  Total cells in the cross-tabulation of all variables \\ 
	$y_1,y_2,...,y_k$ & Counts of original data in each cell of the cross-tabulation \\ 
	$s_1,s_2,...,s_k$ & Counts of synthetic data in each cell of the cross-tabulation \\
	$100~\Sigma_{i = 1,k}({y_i = 1})/N$ & Percentage of unique records in the original data ($p1$) \\
	$100~\Sigma_{i = 1,k}({y_i = 1)}/k$ & Percentage of empty cells in the cross-tabulation ($p0$)\\
	$100~\Sigma_{i = 1,k}({s_i = 1~ \&~ {y_i = 1}})/k~~$   & Percentage of  uniques in the synthetic data that are \\
	& also unique in the original (replicated uniques $ru$)
\end{tabular}

\end{table}

	\subsection{Utility}

Raab et al.\cite{utility2022} have carried out an extensive review of utility measures that have been proposed for synthetic data. The measures are computed by first attempting to discriminate the synthetic data from the original according to some method. All the measures evaluated were found to be highly correlated with each other, and in some cases even identical. The model used to discriminate between the real and synthetic data is more important than the choice of measure. Some of the measures have expected values that can be calculated, or obtained by simulation, when the generative model is the correct one that could have produced the original data. The utility measure that we will use here is the propensity score mean-square-error ($pMSE$) whose expected value for a correct generative model is discussed in \cite{snokerss}. A standardised version can be calculated as the ratio of $pMSE$ to its expectation ($S\_pMSE$). For non-DP synthesis this value will have an expected value of 1.0 when the generative model is correct, with higher values indicating poor utility. Experience of using this measure for many synthetic data sets suggests that synthetic data with values of $S\_pMSE$ below 10 provide useful results that agree well with analysis of the original data, and those under 30 generally produce usable results.   

When the model used to evaluate utility consists of a tabulation, the $pMSE$ has a simple form\cite{utility2022} that is identical to the utility measure proposed by Voas and Williams \cite{vw}. The formula for this and its standardised version for  table $m$,  designated as $U_m$, is shown below.   These formulae assume that a synthetic data set of the same size as the original has been produced. A summary measure for the whole data set can be obtained by averaging $U_m$ over a set of M marginals to give $U_M$. When the synthetic data are created by a method like $ipf$ that constrains a set of marginals, then the expected value of $U_m$ for each constrained marginal will be 1.0. 

	\begin{table}[ht]
	\begin{tabular}{lll}
		$M$ & Number of margins selected \\
		$m_1,m_2,...m_M$ & Number of cells in the $mth$ margin \\
		$yi_1,yi_2,...,yi_m$ & Counts of original data in each cell of the $ith$ margin \\ 
		$si_1,si_2,...,si_m$ & Counts of synthetic data in each cell of the $ith$ margin \\
		$pMSE_m = \Sigma_j{[(yi_j - si_j  )^2/{(yi_j + si_j)/2}]}~~$ &  $pMSE$ for the $ith$ margin \\
		$df_m$ & Degrees of freedom for the $mth$ marginal \\
		$U_m = pMSE_m/df_m$ & $S\_pMSE$ standardised utility for the  $mth$ marginal \\
		$U_M = \Sigma_m{U_m}/M$ & Average standardised utility for the set of $M$ marginals.\\
	\end{tabular}
	
\end{table}
\section{Data sets used for the evaluation \label{data}}
\begin{table}[ht]
	\centering
	\begin{tabular}{lrrrrrrrrrrrr}
		\hline
		& S3 & S5 & S7 &~~~~~&  P3 & P5 & P7 & P7x &~~~~~&  Ps3 & Ps5 & Ps7 \\ 
		\hline
		\textit{N} &  \multicolumn{3}{c}{5,000} & & \multicolumn{4}{c}{1,035,201}  & & \multicolumn{3}{c}{13,309}  \\ 
		\textit{k}& 60 & 2,160 &  77,760 & &  70 & 2,100 & 94,500 & 3,420,900 &  & 70 & 2,100 & 94,500 \\ 
		$p0$ &   0 & 6.23 &  98.20 & &  5.71 & 37.62 & 94.49 & 96.62 & & 34.29 & 73.19 & 99.10 \\ 
		$p1$ &   0 & 6.74 & 35.84 & & 0.00 & 0.02 & 0.12 & 4.51 & & 0.03 & 0.96 & 2.66 \\
		
		\hline
		\\
	\end{tabular}
	\caption{Features of the data sets.} 
	\label{table:dsets}
\end{table}

Data sets used to evaluate the DP synthesis methods were subsets of variables from two sources.  The data set (SD2011) is a sample of 5000 records from a survey on the quality of life in Poland in 2011\footnote{available as a data set as in the \textit{synth}\textbf{\textit{pop}} package}. In all cases any numeric variables were grouped into 5 classes\footnote{The only numeric variables selected from SD2011 were Age and Income, and from the PUMS data only Age.}.
 The data used to synthesise was a subset of the first 3, first 5 and first 7 variables in this data set, labelled S3, S5 and S7. 
  The second  data set was supplied to teams from National Statistics Agencies (NSOs) as part of the evaluation of a guide to synthetic data for NSOs \cite{guide}.  It consisted of an extract of just over a million records from the American Community Survey, made available by the IPUMS project\footnote{https://www.ipums.org/}. As well as demographic data it included an area identifier that divided the USA into 181 Public Use Microdata Areas (PUMAs). Data extracts from this source were 3, 5 and 7 demographic variables (P3, P5 and P7), then one with 7 variables including PUMA (P7x) and finally three smaller data sets from one PUMA with 13 thousand records (Ps3, Ps5 and Ps7). This is important since better utility can be obtained for this type of data by stratifying it into subsets by area. If each subset is made DP the overall $\epsilon$ is the maximum of those used for each subset, since a person can only be in one PUMA. Features of the data sets are summarised in Table \ref{table:dsets} and the variables in each are given in the Appendix \ref{vars}.
  
  The number of cells in the cross-tabulation of all the variables ranged from 60 to over 3 million. Extracts with 7 variables had the most sparse tables (high $p0$). The \% of unique records is a disclosure measure for the original data that is an upper limit for $ru$ from synthetic data sets.  For a given sample size $p0$ and $p1$ increase with the number of cells in the table and they are greatest for small sample sizes. Two of the three data sets with just three variables have no unique records, and the third has only a very small number of unique records. Those datasets with 7 variables are all very sparse with two having over 98\% of the cells in the cross-tabulation as zeros.

\section{Results}
\subsection{Utility and disclosure risk for non-DP synthesis}

Each of the 10 datasets was synthesised using $synthpop$ by the method $catall$ and then by $ipf$ with all two-margins. Results  for disclosiveness and utility, averaged over 10 replications, are shown in Table \ref{table:nonDP}. Synthesis by $catall$ reduces the disclosure risk, as assessed by $ru$, to an average of about 37\% of $p1$ and to a greater extent (varying by data set) for $ipf$ synthesised from two-way margins. As expected all the two-way and three-way utility measures are centred on 1.0 for $catall$. This is also true for synthesis by  $ipf$ for utility evaluated from two-way margins, but for utility evaluated from three-way margins we find the expected lack-of-fit from the incompatible generating model. There was very little evidence of three-way interactions in the SD2011 data sets, but more for the PUMS data. For synthesis by $ipf$ we show the utility for the three-way marginal that gave the worst utility in an evaluation of a two-way synthesis.  This choice of worst marginal was carried forward into the DP syntheses reported below. This will allow the loss of utility due to an incompatible margin to be compared to that due to the addition of DP noise.  For the PUMS data the variables with the most evidence of a three way interaction were age and sex with either marital status or group quarters.

\begin{table}[ht]
	\centering
	\begin{tabular}{lrrrrrrrrrrrr}
		\hline
	& S3 & S5 & S7 &~~~~~& P3 & P5 & P7 & P7x &~~~~~& Ps3 & Ps5 & Ps7 \\ 
	\hline
	\\
	\multicolumn{10}{l}{Original data}\\
	$p1$ & 0.00 & 6.74 & 35.84 &~~~~~& 0.00 & 0.02 & 0.12 & 4.51 &~~~~~& 0.03 & 0.96 & 2.66 \\
	\hline
	\\
	\multicolumn{10}{l}{\textit{catall} saturated model}  & \\
	\% replicated uniques $ru$ &  & 2.44 & 13.16 &~~~~~&  & 0.01 & 0.04 & 1.66 &~~~~~& 0.01 & 0.34 & 0.95 \\ 
	 ($ru$) as a \% of $p1$ &  & 36 & 37 &~~~~~&    & 37 & 36  &   37  &~~~~~& 28 & 36 & 36 \\ \\
	Mean two-way utility & 0.97 & 1.02 & 1.00 &~~~~~& 0.93 & 1.01 & 0.92 & 0.99 &~~~~~& 1.10 & 1.02 & 0.99 \\ 
	Mean.three-way.utility & 0.92 & 1.03 & 0.99 &~~~~~& 0.96 & 0.98 & 0.97 & 1.02 &~~~~~& 1.05 & 1.01 & 1.00 \\
	\hline \\
	\multicolumn{10}{l}{\textit{ipf} with all two-way margins} & \\
	\% replicated.uniques($ru$) &  & 1.67 & 6.41 &~~~~~&  & 0.00 & 0.03 & 1.00 &~~~~~& 0.00 & 0.28 & 0.63 \\ 
		 ($ru$) as a \% of $p1$ &  & 30 & 13 &~~~~~&    & 25 & 17  &   22  &~~~~~& 15 & 22 & 24 \\ \\
	Mean two-way utility & 0.97 & 0.96 & 1.04 &~~~~~& 1.09 & 0.98 & 0.98 & 1.04 &~~~~~& 0.97 & 0.98 & 1.07 \\ 
	Mean three-way utility & 1.48 & 1.76 & 1.77 &~~~~~& 4.41 & 20.58 & 10.34 & 10.26 &~~~~~& 1.54 & 2.14 & 1.67 \\ 
	Worst.three-way.utility & 1.48 & 1.92 & 2.37 &~~~~~& 4.41 & 81.74 & 81.40 & 82.14 &~~~~~& 1.54 & 5.52 & 1.20 \\ 
	\hline
		\\
	\end{tabular}

	\caption{Non DP synthesis results for the 10 data sets, all results are average for 10 syntheses.\label{table:nonDP}} 
\end{table}

\subsection{Utility and disclosure risk for DP synthesis}
\begin{figure}[ht]
	\centering
	\caption{Percentage of replicated uniques for $ipf$ DP and non-DP syntheses expressed as a \% of the percentage of uniques in the original data set. Results for each of 7 data sets with DP method of adding Laplace noise to the two-way margins.}
	\includegraphics[width=1\textwidth]{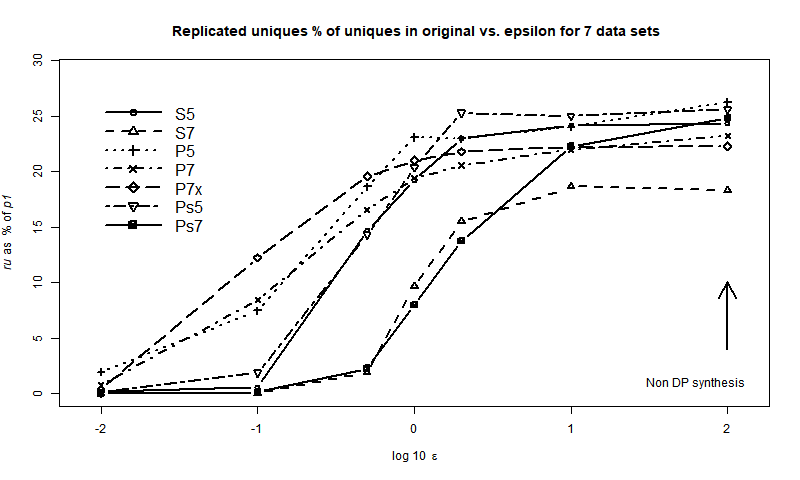}
	\label{fig:1}
\end{figure}

The results for DP synthesis by $catall$ confirmed the expectation that this method would not prove useful; similar results were found by Bowen and Liu \cite{BowenLiu2020}. Even for some larger values of $\epsilon$ utilities often exceeding 100 times their value for non-DP synthesis. The disclosure, measured by $ru$ did fall to low levels, especially for data sets with very sparse cross-tabulations, but this was not associated with acceptable utility for any of the data sets.  The process of adding noise leads to many negative values.  These have to be set to either zero or a small value to allow the probabilities to be used in generating the synthetic data. Exactly how steps 5 and 6 distort the data for these sparse tables depends on the relative proportion of 0s and 1s in the cross-tabulation. For small $\epsilon$ applied to sparse tables step 5 increases the sum of the noisy counts and step 6 reduces large counts. The US Census Bureau have noted the contribution of this type of post-processing to their area counts \cite{Hawes2020,Garfinkel}, but the top-down solution they have used \cite{tda}, including some agreed fixed totals, does not appear applicable to synthetic data generation. Post-processing by steps 5 and 6 are less of a problem for $ipf$ because counts in the margins are larger. Some teams using marginal approaches in the NIST challenge appear to have used the method described here\cite{Zhang2014}, but better methods could be explored. 

\begin{table}[ht]
	\centering
	\begin{tabular}{lrrrrrrrrrr}
		\hline
		
		& S3 & S5 & S7 & P3 & P5 & P7 & P7x & Ps3 & Ps5 & Ps7 \\ 
		\hline
		\multicolumn{5}{l}{$\epsilon$~~~~0.01}\\
		\hline
		\% $ru$ & 0.00 & 0.01* & 0.00* & 0.00 & 0.00 & 0.00 & 0.02 & 0.00 & 0.00 & 0.00 \\ 
$ru$ as a \% of $p1$ &  & 0.18 & 0.01 &  & 1.95 & 0.75 & 0.41 & 0.00 & 0.08 & 0.00 \\ 
Mean two-way utility & 245.71 & 325.45 & 365.47 & 83.03 & 412.38 & 1942.99 & 1692.89 & 227.55 & 767.38 & 1240.11 \\ 
Mean three-way utility & 139.81 & 135.90 & 133.32 & 62.22 & 258.10 & 949.60 & 786.60 & 131.96 & 365.65 & 543.98 \\ 
worst three-way utility & 139.81 & 221.15 & 159.31 & 62.22 & 501.79 & 1236.90 & 1211.84 & 131.96 & 501.07 & 441.72 \\ 
\hline
		\multicolumn{5}{l}{$\epsilon$~~~~0.1}\\
\hline
\% $ru$ & 0.00 & 0.04 & 0.02* & 0.00 & 0.00 & 0.01 & 0.55 & 0.00 & 0.02 & 0.00 \\ 
$ru$ as a \% of $p1$ &  & 0.56 & 0.05 &  & 7.46 & 8.45 & 12.24 & 0.00 & 1.88 & 0.14 \\  
Mean two-way utility & 12.63 & 99.01 & 172.29 & 5.47 & 23.05 & 128.11 & 104.59 & 16.48 & 81.11 & 261.94 \\ 
Mean three-way utility & 10.27 & 56.77 & 79.06 & 9.51 & 32.04 & 78.55 & 61.82 & 11.46 & 49.17 & 135.13 \\ 
worst three-way utility & 10.27 & 75.29 & 94.35 & 9.51 & 102.79 & 136.49 & 159.38 & 11.46 & 58.53 & 149.41 \\ 
\hline
\multicolumn{5}{l}{$\epsilon$~~~~0.5}\\
\hline
		\% $ru$ & 0.00 & 0.99 & 0.68* & 0.00 & 0.00 & 0.02 & 0.88 & 0.00 & 0.14 & 0.06 \\ 
		$ru$ as a \% of $p1$ &  & 14.69 & 1.90 &  & 18.64 & 16.58 & 19.57 & 10.00 & 14.30 & 2.26 \\
		Mean two-way utility & 1.69 & 14.59 & 31.67 & 1.38 & 3.51 & 19.66 & 15.65 & 2.85 & 15.16 & 46.60 \\ 
		Mean three-way utility & 1.90 & 8.91 & 20.85 & 4.61 & 21.67 & 21.20 & 17.45 & 2.89 & 9.82 & 28.59 \\ 
		worst three-way utility & 1.90 & 9.70 & 24.30 & 4.61 & 81.96 & 91.95 & 92.17 & 2.89 & 18.52 & 7.38 \\ 
		\hline
		\multicolumn{5}{l}{$\epsilon$~~~~1}\\
		\hline
		\% $ru$ & 0.00 & 1.29 & 3.46 & 0.00 & 0.00 & 0.02 & 0.95 & 0.00 & 0.20 & 0.21 \\ 
		$ru$ as a \% of $p1$ &  & 19.20 & 9.66 &  & 23.08 & 19.41 & 20.99 & 10.00 & 20.39 & 7.99 \\ 
		Mean two-way utility & 1.15 & 5.48 & 15.21 & 1.14 & 2.04 & 8.32 & 6.58 & 2.09 & 5.34 & 22.03 \\ 
		Mean three-way utility & 1.40 & 4.21 & 9.51 & 4.34 & 20.72 & 14.47 & 12.77 & 2.06 & 4.69 & 14.50 \\ 
		worst three-way utility & 1.40 & 4.89 & 11.82 & 4.34 & 80.48 & 84.96 & 82.32 & 2.06 & 9.18 & 2.78 \\ 
		\hline
		\multicolumn{5}{l}{$\epsilon$~~~~2}\\
		\hline
		\% $ru$ & 0.00 & 1.55 & 5.55 & 0.00 & 0.00 & 0.02 & 0.98 & 0.01 & 0.24 & 0.37 \\ 
		$ru$ as a \% of $p1$ &  & 23.03 & 15.49 &  & 22.96 & 20.57 & 21.79 & 17.50 & 25.31 & 13.76 \\ 
		Mean two-way utility & 1.08 & 2.84 & 5.86 & 0.85 & 1.29 & 3.90 & 3.32 & 1.36 & 3.31 & 11.86 \\ 
		Mean three-way utility & 1.31 & 2.65 & 4.37 & 3.97 & 20.27 & 11.87 & 11.16 & 1.65 & 3.27 & 8.16 \\ 
		worst three-way utility & 1.31 & 3.16 & 5.57 & 3.97 & 79.48 & 82.24 & 80.19 & 1.65 & 7.10 & 1.66 \\ 
		\hline
		\multicolumn{5}{l}{$\epsilon$~~~~10}\\
		\hline
		\% $ru$ & 0.00 & 1.63 & 6.68 & 0.00 & 0.00 & 0.03 & 1.00 & 0.01 & 0.24 & 0.59 \\ 
		$ru$ as a \% of $p1$ &  & 24.15 & 18.64 &  & 24.02 & 21.96 & 22.18 & 35.00 & 25.00 & 22.26 \\ 
		Mean two-way utility & 1.15 & 1.15 & 1.64 & 0.91 & 1.07 & 1.32 & 1.54 & 1.05 & 1.25 & 2.55 \\ 
		Mean three-way utility & 1.40 & 1.82 & 2.05 & 4.06 & 20.02 & 10.38 & 10.20 & 1.52 & 2.28 & 2.58 \\ 
		worst three-way utility & 1.40 & 2.14 & 2.53 & 4.06 & 78.29 & 80.65 & 80.81 & 1.52 & 5.97 & 1.26 \\ 
		\hline
		\\
	\end{tabular}
	\caption{Disclosure measures and utility for differentially private $ipf$ synthesis with 6 values of $\epsilon$ and 10 data sets. All figures are averages over 10 independent syntheses. Rows marked with * indicate that one or more of the $ipf$ fits failed to converge in 5,000 iterations\label{table:ipf}}
		
	\end{table}

Table \ref{table:ipf} displays the disclosure risk and utility for the 10 data sets and six different choices of $\epsilon$ synthsised by $ipf$. In a few cases with small values of $\epsilon$ the fitting procedure failed to converge in 5,000 iterations.\footnote{It is possible that the added noise produced margins that are impossible to reconcile. These cases do not correspond to useable syntheses.}   Figure \ref{fig:1} plots $ru$ as a \% of $p1$ against $log_{10}({\epsilon})$, with the non-DP synthesis being plotted at 3, equivalent to an $\epsilon$ of 100. Disclosure risk, as measured by replicated uniques decreases with $\epsilon$ as we would expect. 
Two data sets (S7 and Ps7) each with 7 variables and a relatively small sample size, show a good reduction in $ru$ by an $\epsilon$ of 1, and  better for lower values. To evaluate utility, we first consider two-way margins that will not be affected by the lack-of-fit found for three-way margins in the non-DP synthesis. For the data sets analysed here $\epsilon$ values of 1 and above give satisfactory utility, judged by having average values below 30. Other data sets require a lower $\epsilon$ of 0.1 to get an improvement in $\%ru$, and hence higher (worse) utility measures. 
 An $\epsilon$ of 0.5  gives satisfactory utility for most of the data sets. For the lower values of $\epsilon$ used here (0.1 and 0.01)  utility is too poor to be acceptable. Additional exploratory checks illustrated the failure of the synthetic data to show the same relationships between variables as was found in the original data. The process of adding noise to the margins has the effect of bringing them all closer to a constant value and thus inevitably weakening the relationships found in the original. Looking at the results for utility from three-way tables, we can see that the results are limited by the utility that can be achieved by non-DP synthesis with the same models. In many cases, especially for the worst three-way interaction this lack-of-fit to the generative model is more damaging to utility than the DP adjustment. 
 All of the evaluations were carried out with a very small value of a prior $\alpha$ added to each cell of the tables\footnote{A count determined by the parameter $nprior$ is distributed equally over the table or margin entries except those defined as structural zeros. This is important for non-DP synthesis so as to prevent them remaining as zeros. The default value of 1 for $nprior$ was used.}. Increasing $\alpha$ can itself contribute to preventing disclosure  \cite{jackson_et_al2021}, as used in the Dirchilet-Multinomial synthesiser \cite{onthemap}. Using this approach without any DP adjustment gave some improvement in $ru$ for large $\alpha$ but at the expense of greatly damaged utility. When $\alpha$ was adjusted for DP syntheses, its influence was much less than  that of the DP parameter.

\section{Discussion and future work}
We have introduced a very simple modification of two methods to allow them to produce DP synthetic data. The saturated model is useful  for non-DP data synthesis, well suited to large administrative data bases of count data. But efforts, so far, to make this model differentially private find  poor utility, even for large values of $\epsilon$ that provide little improvement in disclosure.

In contrast, the $ipf$ method, based on margins, seems to be easily adapted to provide DP synthetic data. This model can be fitted in $synthpop$ with  code like this example. This code creates a synthetic data object (\texttt{synipf}) by the default method of fitting all two-way interactions. The percentage of replicated uniques can then be found with the next line, and the third line calculates the average two-way utility for all two-way interactions. The default utility measure,  $pMSE$, is used but there is a choice of a further 15 measures available.

\begin{verbatim}
	 synipf <- syn(S7, method= "ipf", ipf.priorn = 0, ipf.epsilon = 1)
	  
	 ru_S7_1 <- replicated.uniques(synipf, S7)$per.replications
	 
	 util_S7_2way <- mean(utility.tables(synipf,SD20113,"twoway", plot = FALSE)$tabs[,2])
 
\end{verbatim}

This is just a preliminary version of DP synthesis with $synthpop$.
It has some clear limitations. In particular, it is limited to data sets with a relatively small number of variables because of the need to store the complete cross tabulation. The teams using marginal models for the NIST challenge have used graphical models to define the parameters of the models.  It may be possible to make their open-source routines, usually written in $Python$, accessible within R.  A further limitation is that the methods apply only to categorical variables. Non-DP versions of $catall$ and $ipf$ can be used for numeric data by asking the program to categorise the variables, and then select from the groups in the original data at the end of the synthesis. For DP synthesis a method that did not access the original data again would be required.  
Many other modifications could be attempted. In particular options should be made available to add noise to the margins by methods other than a Laplace distribution. Examples are trimmed $\epsilon-\delta$ Gausian noise and the Exponential mechanism, see \cite{mcKwinning} for other choices. 

Further investigation of the disclosure risks posed by synthetic data, both  DP and non-DP, would be helpful. These should include realistic evaluations of the behaviour of those attempting to find confidential information about data subjects. A disclosure from synthetic data that turns out to be false can also cause harm, both to the data subject and to the reputation of the data holder. This underlines the importance of ensuring that everyone who has access to synthetic data is aware that it is not real.

Inference from synthetic data will always be limited by the model used in creating it. George Box's caveat that ``all models are wrong, but some are useful" needs to be borne in mind when using results from synthetic data whether DP or not.  There is a powerful argument that no important decisions should be taken using analyses of synthetic data. Confirmatory analyses on the real data, perhaps via a validation server, should be carried out. But synthetic data can still have an important role in widening access to confidential data and in providing realistic data sets for training.
\section{Acknowledgements}
This work would not have been possible without the work of Beata Nowok, the main author of \textit{synth}\textbf{\textit{pop}}. Any errors found in the new DP routines on github (see note 2) are entirely my responsibility. The ESRC/UKRI provided support for the Administrative data Research Centre and the Scottish Longitudinal Study. I would also like to thank two anonymous referees for helpful comments on an earlier version of this paper.

\bibliographystyle{splncs03}
\bibliography{psd2022_raab}

\section{Appendix}
\subsection{Details of the variables in data sets \label{vars}}
Tables \ref{table:t4} and \ref{table:t5} give details of the variables selected from each of the two data sets. See section \textbf{data} for how each of the data set were created. The two Age variables were each grouped into 5 categories.

\begin{table}[ht]
	\centering
	\begin{tabular}{rlrr}
		\hline
		& Variable & Number of missing values & Number of distinct values \\ 
		\hline
		1 & Sex &   0 &   2 \\ 
		2 & Age grouped &   0 &  5 \\ 
		3 & Placesize &   0 &   6 \\ 
		4 & Education level &   7 &  4  \\ 
		5 & Social and professional group &   33 &   9 \\ 
		6 & Income grouped  & 683, 603 not applicable &  5 \\ 
		7 & Marital status &  9 &   6 \\ 
		\hline
		\\
			\end{tabular}
		\caption{Variables seleced from the SD2011 data set.}
\label{table:t4}
\end{table}

\begin{table}[ht]
	\centering
	\begin{tabular}{rlrr}
		\hline
		& Variable & Number of missing values & Number of distinct values \\ 
		\hline
		1 & Public Use Microdata Area &   0 & 181 \\ 
		2 & Year &   0 &   7 \\ 
		3 & Group Quarters &   0 &   5 \\ 
		4 & Sex &   0 &   2 \\ 
		5 & Age &   0 &  73 \\ 
		6 & Marital Status &   0 &   6 \\ 
		7 & Race &   0 &   9 \\ 
		8 & Hispanic &   0 &   5 \\ 
		\hline
		\\
			\end{tabular}
			\caption{Variables seleced from the IPUMS data set.}
	\label{table:t5}
\end{table}
\end{document}